\documentclass[iop]{emulateapj}
\usepackage{amsmath}
\usepackage{subfigure}
\usepackage{graphicx}

\graphicspath{{figs/}}

\newcommand{\pd}[2]{ \frac{\partial #1}{\partial #2} }

\newcommand{\vsp}[1]{ \mathbf #1 }
\newcommand{\Pm}{\rm{Pm}}

\newcommand{\figext}[1]{#1.pdf}

\makeatletter
\def\ScaleIfNeeded{%
  \ifdim\Gin@nat@width>\linewidth
  \linewidth
  \else
  \Gin@nat@width
  \fi
}
\makeatother

\shorttitle{BNS Merger Dynamo}
\shortauthors{J. Zrake and A. I. MacFadyen}
\begin{document}

\title{Magnetic energy production by turbulence in binary neutron star mergers}
\author{Jonathan Zrake and Andrew I. MacFadyen}

\affil{Center for Cosmology and Particle Physics, Physics Department,
  New York University, New York, NY 10003, USA}

\begin{abstract}

The simultaneous detection of electromagnetic and gravitational wave
emission from merging neutron star binaries would aid greatly in their
discovery and interpretation. By studying turbulent amplification of
magnetic fields in local high-resolution simulations of neutron star
merger conditions, we demonstrate that magnetar-level ($\gtrsim
10^{16}$G) fields are present throughout the merger duration. We find
that the small-scale turbulent dynamo converts 60\% of the randomized
kinetic energy into magnetic fields on a merger time scale. Since
turbulent magnetic energy dissipates through reconnection events which
accelerate relativistic electrons, turbulence may facilitate the
conversion of orbital kinetic energy into radiation. If $10^{-4}$ of the
$\sim 10^{53}$ erg of orbital kinetic available gets processed through
reconnection, and creates radiation in the 15-150 keV band, then the
fluence at 200 Mpc would be $10^{-7} \rm{erg}/\rm{cm^2}$, potentially
rendering most merging neutron stars in the advanced LIGO and Virgo
detection volumes detectable by \emph{Swift} BAT.

\end{abstract}

\maketitle

\section{Introduction}

The in-spiral and coalescence of binary neutron star systems is a topic
of increasingly intensive research in observational and theoretical
astrophysics. It is anticipated that the first direct detections of
gravitational wave (GW) will be from compact binary mergers. Binary
neutron star (BNS) mergers are also thought to produce short-hard
gamma-ray bursts (SGRB's) \citep{Eichler:1989p126, Narayan:1992p83,
  Belczynski:2006p1110, Metzger:2008p5119}. Simultaneous detections of a
prompt gravitational wave signal with a spatially coincident
electromagnetic (EM) counterpart dramatically increases the potential
science return of the discovery. For this reason, there has been
considerable interest as to which, if any, detectable EM signature may
result from the merger \citep{Metzger:2012p4995, Piran:2013p771}. Other
than SGRBs and their afterglows, including those viewed off-axis
\citep{Rhoads:1997p1, vanEerten:2011p37}, suggestions include optical
afterglows associated with the radio-active decay of tidally expelled
r-process material\citep{Li:1998p5004, Metzger:2010p5000} (though
detailed calculations indicate they are faint \citep{Barnes:2013p346}),
radio afterglows following the interaction of a mildly relativistic
shell with the interstellar medium \citep{Nakar:2011p5001}, and
high-energy pre-merger emission from resistive magnetosphere
interactions \citep{Hansen:2001p4998, Piro:2012p4994}.

Merging neutron stars possess abundant orbital kinetic energy ($\sim
10^{53}$ergs). A fraction of this energy is certain to be channelled
through a turbulent cascade triggered by hydrodynamical instabilities
during merger. Turbulence is known to amplify magnetic fields by
stretching and folding embedded field lines in a process known as the
small-scale turbulent dynamo \citep{Vainshtein:1972p5054,
  Chertkov:1999p4065, Tobias:2011p4407,
  Beresnyak:2012p5042}. Amplification stops when the magnetic energy
grows to equipartition with the energy containing turbulent eddies
\citep{Schekochihin:2002p84, Schekochihin:2004p4973,
  Federrath:2011p5044}. An order of magnitude estimate of the magnetic
energy available at saturation of the dynamo can be informed by global
merger simulations. These studies indicate the presence of turbulence
following the nonlinear saturation of the Kelvin-Helmholtz (KH)
instability activated by shearing at the NS surface layers
\citep{Price:2006p3270, Liu:2008p4899, Anderson:2008p4898,
  Rezzolla:2011p4950, Giacomazzo:2011p4902}. The largest eddies produced
are on the $\sim 1$ km scale and rotate at $\sim 0.1c$, setting the
cascade time $t_\mathrm{eddy}$ and kinetic energy injection rate
$\varepsilon_K \equiv E_K/t_\mathrm{eddy}$ at $0.1$ms and $5 \times
10^{50} \mathrm{erg/s}$ respectively. When kinetic equipartition is
reached, each turbulent eddy contains $5 \times 10^{46} \mathrm{erg}$ of
magnetic energy, and a mean magnetic field strength
\begin{equation}
  B_\mathrm{RMS} \gtrsim 10^{16} \mathrm{G}
  \left(\frac{\rho}{10^{13} \mathrm{g/cm^3}} \right)^{1/2}
  \left(\frac{v_\mathrm{eddy}}{0.1c} \right).
\end{equation}

Whether such conditions are realized in merging neutron star systems
depends upon the dynamo saturation time $t_\mathrm{sat}$ and
equipartition level $E_\mathrm{sat}/E_K$. In particular, if
$t_\mathrm{sat} \lesssim t_\mathrm{merge}$ then turbulent volumes of
neutron star material will contain magnetar-level fields throughout the
early merger phase. Once saturation is reached, a substantial fraction
of the injected kinetic energy, $0.7 \varepsilon_K$, is resistively
dissipated \citep{Haugen:2004p5062} at small scales. Magnetic energy
dissipated by reconnection in optically thin surface layers will
accelerate relativistic electrons \citep{Blandford:1987p4,
  Lyutikov:2003p893}, potentially yielding an observable electromagnetic
counterpart, independently of whether the merger eventually forms a
relativistic outflow capable of powering a short gamma-ray burst.

In this Letter we demonstrate that the small-scale turbulent dynamo
saturates quickly, on a time $t_\mathrm{sat} \lesssim t_\mathrm{merge}$,
and that $\gtrsim 10^{16}$G magnetic fields are present throughout the
early merger phase. This implies that the magnetic energy budget of
merging binary neutron stars is controlled by the rate with which
hydrodynamical instabilities randomize the orbital kinetic energy. Our
results are derived from simulations of the small scale turbulent dynamo
operating in the high-density, trans-relativistic, and highly conductive
material present in merging neutron stars. We have carefully examined
the approach to numerical convergence and report grid resolution
criteria sufficient to resolve aspects of the small-scale dynamo. Our
Letter is organized as follows. The numerical setup is briefly described
in Section 2. Section 3 reports the resolution criterion for numerical
convergence of the dynamo completion time and the saturated field
strength. In Section 4 we asses the possibility that magnetic
reconnection events may convert a sufficiently large fraction of the
magnetic energy into high energy photons to yield a prompt
electromagnetic counterpart detectable by high energy observatories
including \emph{Swift} and \emph{Fermi}.

\begin{figure}
  \centering \includegraphics[width=3.4in]{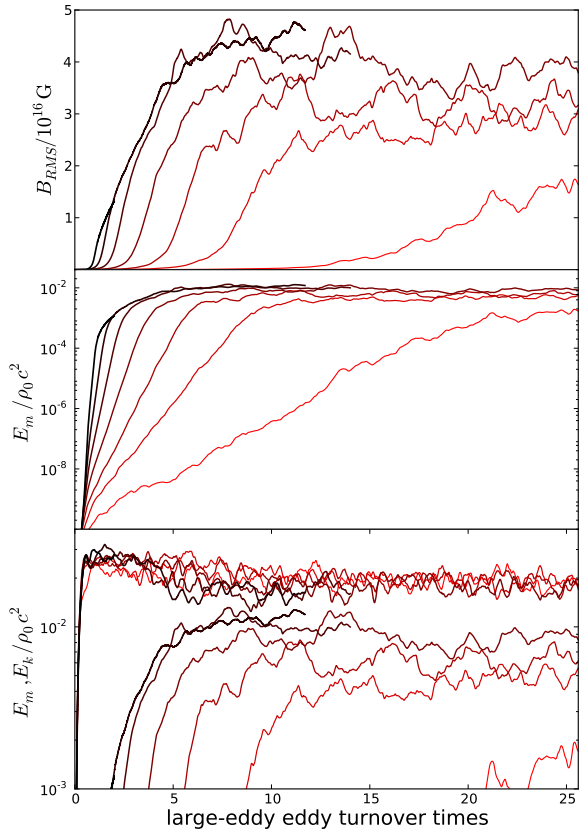}
  \caption{Time development of volume-averaged kinetic and magnetic
    energies at resolutions between $16^3$ and $1024^3$. Lower
    resolutions are shown in red and graduate to black with higher
    resolution. \emph{Top}: The root mean square magnetic field strength
    in units of $10^{16}\rm{G}$. When a turbulent volume is resolved by
    $16^3$ zones, the small-scale dynamo proceeds so slowly that almost
    no amplification is observed in the first 1ms. \emph{Middle}: The
    magnetic energy in units of the rest mass $\rho_o c^2$ shown on
    logarithmic axes. It is clear that the linear growth rate increases
    at each resolution. \emph{Bottom}: The kinetic energy (upper curves)
    shown again the magnetic energy (lower curves) again in units of
    $\rho_o c^2$. For all resolutions, the kinetic energy saturates in
    less than 1 $t_\mathrm{eddy}$.}
  \label{fig:energy-growth}
\end{figure}

\begin{figure*}
  \centering \includegraphics[width=6.5in]{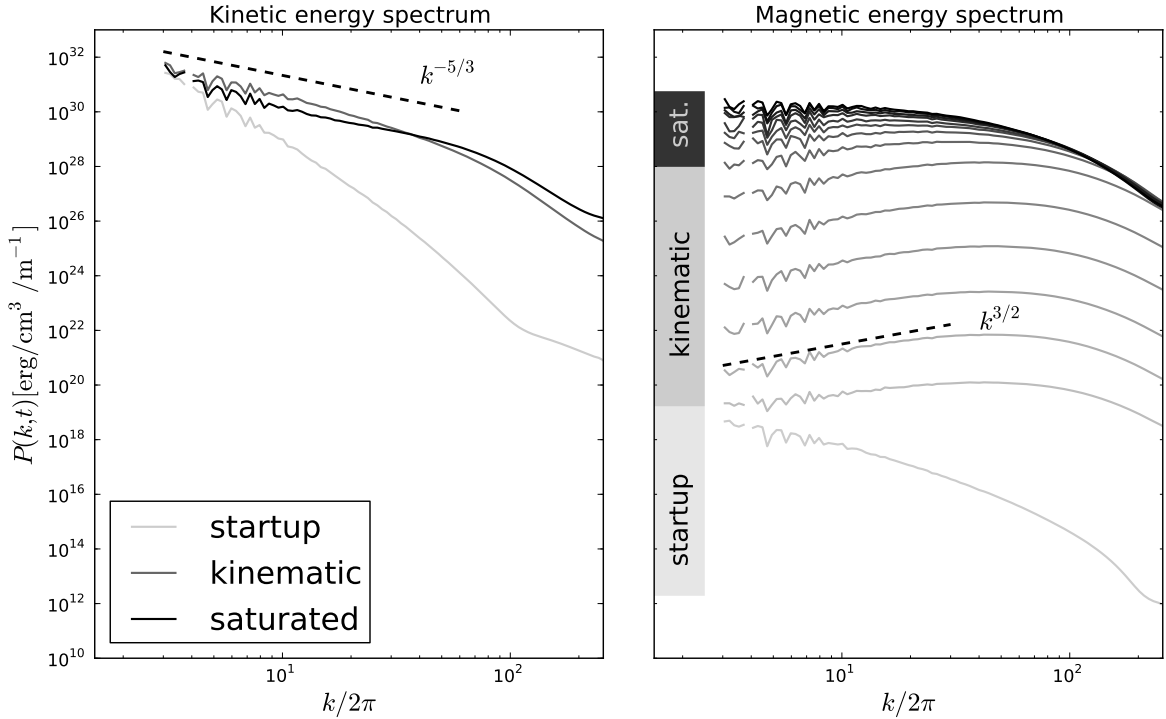}
  \caption{Time development of the spectrum of kinetic (left) and
    magnetic (right) energy, given in $\rm{erg}/cm^3/m^{-1}$ for
    resolution $512^3$. During the \emph{startup} phase the hydrodynamic
    cascade is not yet fully developed, this phase lasts for $1
    t_\mathrm{eddy}$. The \emph{kinematic} phase refers to the time when
    the hydrodynamic cascade is fully developed, but the magnetic energy
    is still energetically sub-dominant. During the \emph{kinematic}
    phase, the the kinetic energy power spectrum is consistent with a
    Kolmogorov cascade, having a slope $k^{-5/3}$ over intermediate
    wavelengths. Meanwhile, the magnetic energy spectrum is consistent
    with the Kazentsev description, having a positive slope $k^{3/2}$
    over the same intermediate wavelengths. During this phase the
    magnetic energy exponentiates rapidly, with an e-folding time
    controlled by shearing at the smallest available scales. After
    \emph{saturation}, the magnetic energy spectrum conforms to the
    kinetic energy spectrum, such that $E_M(k) \approx 4 E_K(k)$ over
    the intermediate wavelengths. The kinetic energy spectrum after the
    dynamo completion has a slightly shallower slope than during the
    kinematic phase, but is still consistent with a $5/3$ spectral
    index.}
  \label{fig:pspec-time-devel-res}
\end{figure*}

\section{Numerical scheme and physical setup}

The equations of ideal relativistic magnetohydrodynamics (RMHD) have
been solved on the periodic unit cube with resolutions between $16^3$
and $1024^3$.
\begin{subequations}\label{eqn:rmhd-system}
  \begin{align}
    \nabla_\nu N^\nu &= 0 \\
    \nabla_\nu T^{\mu \nu} &= S^\mu \\
    \pd{\vsp B}{t} &= \vsp \nabla \times (\vsp v \times \vsp B)
  \end{align}
\end{subequations}
Here, $b^\mu = F^\mu_{\ \nu} u^\nu$ is the magnetic field four-vector,
and $h^* = 1+e^*+p^*/\rho$ is the total specific enthalpy, where $p^* =
p_g + b^2/2$ is the total pressure, $p_g$ is the gas pressure and $e^* =
e + b^2/2\rho$ is the specific internal energy. The source term
$S^\mu=\rho a^\mu - \rho (T/T_0)^4 u^\mu$ includes injection of energy
and momentum at the large scales and the subtraction of internal energy
(with parameter $T_0=40$MeV) to permit stationary evolution. Vortical
modes at $k/2\pi \le 3$ are forced by the four-acceleration field $a^\mu
= \frac{du^\mu}{d\tau}$ which smoothly decorrelates over a large-eddy
turnover time, as described in \cite{Zrake:2012p5098, Zrake:2013p12}.

We have employed a realistic micro-physical equation of state (EOS)
appropriate for the conditions of merging neutron stars. It includes
contributions from high-density nucleons according a relativistic mean
field model \citep{Shen:1998p4930, Shen:2010p015806}, a relativistic and
degenerate electron-positron component, neutrino and anti-neutrino pairs
in beta equilibrium with the nucleons, and radiation pressure. For our
conditions, all the components make comparable contributions to the
pressure. We have also employed a simpler gamma-law EOS and found close
agreement for the conditions explored in this paper, indicating that the
MHD effects are insensitive the EOS for trans-sonic conditions. The
models presented in our resolution study use the far less expensive
gamma-law equation of state.

All of the simulations presented in this study use the HLLD approximate
Riemann solver \citep{Mignone:2009p2269}, which has been demonstrated as
crucial in providing the correct spatial distribution of magnetic energy
in MHD turbulence \citep{Beckwith:2011p4448}. The solution is advanced
with an unsplit, second-order MUSCL-Hancock scheme. Spatial
reconstruction is accomplished with the piecewise linear method
configured to yield the smallest possible degree of numerical
dissipation. The divergence constraint on the magnetic field is
maintained to machine precision at cell corners using the finite volume
CT method of \cite{Toth:2002p4505}. Full details of the numerical scheme
may be found in \cite{Zrake:2012p5098}.

\section{Results}
\subsection{Growth and saturation of the magnetic energy}
Magnetic fields are amplified in our simulations by the small-scale
turbulent dynamo. Turbulent fluid motions stretch and fold the magnetic
field lines, causing exponential growth of magnetic energy
\citep[e.g.][]{Moffatt:1978p5105}. This growth is attributed to the
advection and diffusion of $\vsp{B}$ through the MHD induction equation
(Eq. \ref{eqn:rmhd-system}c). When the magnetic field is weak ($E_M \ll
E_K$) $\vsp{B}$ evolves passively, and the turbulence is
hydrodynamical. This limit is referred to as small-scale kinematic
turbulent dynamo, and is well described by Kazentzev's model
\citep{Kazantsev:1968p1031}. This model predicts that the power spectrum
of magnetic energy peaks at the resistive scale $\ell_\eta$ and obeys a
power law $k^{3/2}$ at longer wavelengths. The kinematic phase ends when
the magnetic field acquires sufficient tension to modify the
hydrodynamic motions, after which time a dynamical balance between
kinetic and magnetic energy is established.

Numerical simulations of MHD turbulence are typically limited to
magnetic Prandtl numbers $\Pm \equiv \nu / \eta \approx 1$. However,
neutron star material is characterized by $Pm \gg 1$, with the viscous
cutoff due to neutrino diffusion occurring at around 10 cm, while the
resistive scale is significantly smaller
\citep{Thompson:1993p4921}. However, the disparity between true and
simulated magnetic Prandlt number does not influence our
conclusions. This is because dynamos are generically easier to establish
in the high Pm regime than the small \citep{Haugen:2004p5062,
  Ponty:2011p5045}.

\begin{figure}
  \centering \includegraphics[width=3.4in]{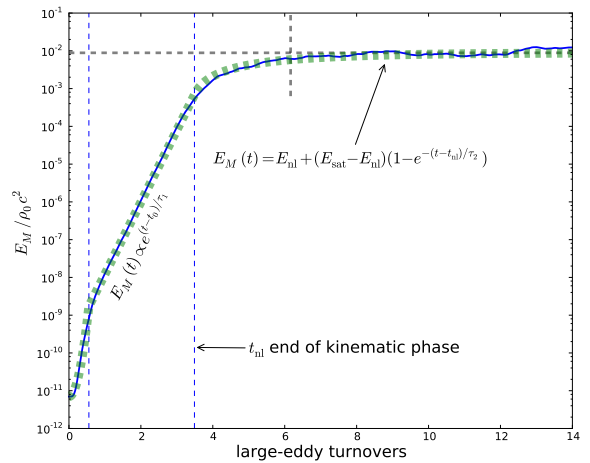}
  \caption{Time history of the magnetic energy for a representative run
    at $128^3$, together with the empirical model (Equation
    \ref{eqn:magnetic-fit}) with best-fit parameters. The horizontal
    dashed line indicates the magnetic energy, $E_\mathrm{sat}$ at the
    dynamo completion. From left to right, the vertical dashed lines
    mark the end of the startup, kinematic, and saturation
    phases.}
  \label{fig:magnetic-fit}
\end{figure}

We use an initially uniform, pulsar-level ($10^{11}$G) seed magnetic
field. This field is sub-dominant to the kinetic energy by 10 orders of
magnitude, so that the initial field amplification is expected to be
well-described by Kazentsev's model. Indeed, we find that during this
phase the power spectrum of magnetic energy follows $k^{3/2}$
(Fig. \ref{fig:pspec-time-devel-res}), peaking at around 10 grid zones,
which we identify as the effective scale of resistivity. The saturation
process begins at ever-earlier times with increasing numerical
resolution. This reflects the fact that during the kinematic phase,
magnetic energy exponentiates on a time scale controlled by shearing at
the smallest scales. In numerically converged runs, full saturation
occurs with $E_M \approx 0.6 E_K$ and is characterized by scale-by-scale
super-equipartition, with $E_K \approx 4 E_M$ at all but the largest
scale.

\subsection{Numerical convergence}
The same driven turbulence model was run through magnetic saturation at
resolutions $16^3$, $32^3$, $64^3$, $128^3$, $256^3$, and
$512^3$. Another model at $1024^3$ was run through the end of the
kinematic phase, but further evolution was computationally
prohibitive. Fig. \ref{fig:energy-growth} shows the development of
$B_{RMS}$, $E_M$, and $E_K$ as a function of time at each resolution.

We find that sufficiently resolved runs ($\ge 512^3$) attain \emph{mean}
magnetic field strengths of $10^{16}$G within two large eddy
rotations. All models with resolutions $\ge 32^3$ eventually attain mean
fields of $\gtrsim 3 \times 10^{16}$G. The saturated field strength
increases until resolution $256^3$. We find that the kinematic growth
rate is higher at each higher resolution, while the time-scale for the
non-linear saturation converges at $256^3$ to roughly five large-eddy
rotation times, or about 0.5ms for the physical parameters of binary
neutron star mergers.

\begin{deluxetable*}{lcc}
  \tablecolumns{3}
  \small
  \tablewidth{0pt}
  \tablecaption{Empirical model for magnetic energy growth}
  \tablehead{\colhead{Fit parameter} & \colhead{Description} &
    \colhead{Numerically converged value}}

  \tablecomments{Model parameters characterizing the growth of magnetic
    energy before and during the dynamo saturation. Numerical
    convergence is attained for the fully saturated magnetic energy
    $E_\mathrm{sat}$ and the dynamo completion time $t_\mathrm{sat}$
    defined as $t_\mathrm{nl} + \tau_2$.}

  \startdata
  $\tau_0$ & Startup time-scale & none, artifact \nl
  $t_1$ & Hydrodynamic cascade fully developed & $t_\mathrm{eddy}$ \nl
  $\tau_1$ & Kinematic growth time-scale & none, $\propto N^{-2/3}$ \nl
  $t_\mathrm{nl}$ & End of kinematic phase & $2t_\mathrm{eddy}$ \nl
  $\tau_2$ & Non-linear saturation time-scale &
  $t_\mathrm{nl} + \tau_2 \approx 5 t_\mathrm{eddy}$, $256^3$ \nl
  %  $E_\mathrm{nl}$ & Magnetic energy at the end of kinematic phase & - \nl
  $E_\mathrm{sat}$ & Saturated magnetic energy & $0.6 E_K$, $512^3$
  \enddata
  \label{tab:magnetic-fit}
\end{deluxetable*}

\begin{figure}
  \centering \includegraphics[width=3.4in]{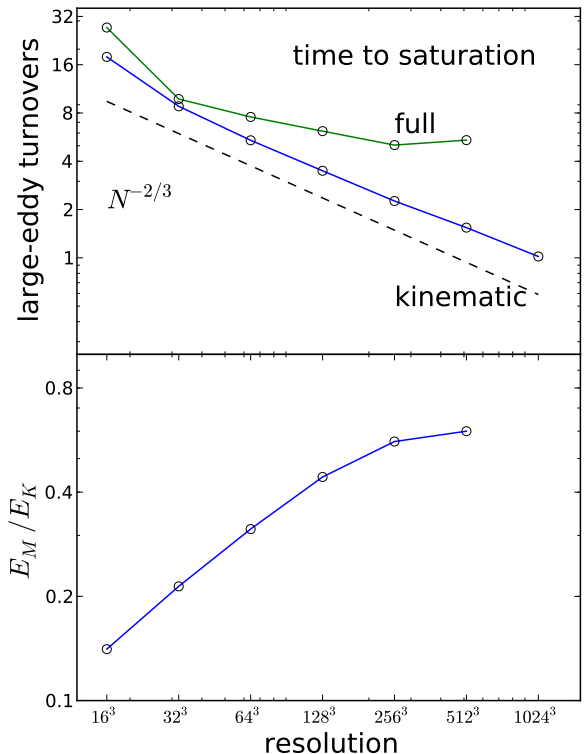}
  \caption{\emph{Top}: Convergence study of the kinematic dynamo growth
    time $\tau_1$ (\emph{blue}) and the dynamo completion time
    $t_\mathrm{sat}$(\emph{green}) defined as $t_\mathrm{nl} +
    \tau_2$. \emph{Bottom}: Convergence study of the best-fit model
    parameter $E_\mathrm{sat}$ expressed as the ratio of magnetic to
    kinetic energy $E_M/E_K$. The converged value of the volume-averaged
    $E_M/E_K \approx 0.6$. Nevertheless, at intermediate wavelengths
    $E_M(k)/E_K(k) \approx 4$. As shown in Figure
    \ref{fig:pspec-time-devel-res}, the largest scales remain
    kinetically dominated. This indicates the suppression of coherent
    magnetic structure formation near the integral scale of turbulence.}
  \label{fig:convergence}
\end{figure}

In order to quantitatively describe the time development of magnetic
energy $E_M(t)$, we describe it with an empirical model,

\begin{equation}
  E_M(t) = \begin{cases} E_M(0) e^{t^2/\tau^2_0} & 0 < t < t_1
    \\ E_M(t_1) e^{(t-t_1)/\tau_1} & t_1 < t < t_\mathrm{nl}
    \\ E_\mathrm{nl} + (E_\mathrm{sat} - E_\mathrm{nl}) (1 -
    e^{-(t-t_\mathrm{nl})/\tau_2}) & t_\mathrm{nl} < t < \infty
  \end{cases}
  \label{eqn:magnetic-fit}
\end{equation}

where the 6 parameters (summarized in Table \ref{tab:magnetic-fit}) are
obtained by a least-squares optimization. Fig. \ref{fig:magnetic-fit}
shows the empirical model given in Equation \ref{eqn:magnetic-fit}
applied to a representative run at $128^3$. The first phase, $t < t_1$
is a startup transient, and lasts until the hydrodynamic cascade is
fully developed at $t_1 \approx t_\mathrm{eddy}$. The kinematic dynamo
phase lasts between $t_1$ and $t_\mathrm{nl}$, during which the magnetic
energy exponentiates on the time scale $\tau_1$. At $t_\mathrm{nl}$, the
smallest scales reach kinetic equipartition and the growth rate
slows. In the final stage, $E_M$ asymptotically approaches
$E_\mathrm{sat}$ on the time-scale $\tau_2$. We define the dynamo
completion time $t_\mathrm{sat}$ as $t_\mathrm{nl} + \tau_2$.
Fig. \ref{fig:convergence} shows the best-fit $E_\mathrm{sat}$,
$t_\mathrm{sat}$, and $\tau_1$ as a function of the resolution.

% Some authors have described the non-linear stage between
% $\t_\mathrm{nl}$ and $t_\mathrm{sat}$ by linear growth of $E_K/E_M$
% \cite{Ryu:2008p909, Beresnyak:2012p5042}. For this study we have found
% it appropriate to use exponential decay toward $E_\mathrm{sat}$,
% having observed that the growth of $E_M$ continues to slow as it
% approaches $E_\mathrm{sat}$.

The magnetic energy $E_\mathrm{sat}$ at dynamo completion shows signs of
converging to a value of $0.6 E_K$ by resolution $512^3$. The time scale
$\tau_2$ on which the magnetic energy asymptotically approaches
$E_\mathrm{sat}$ is consistently $\approx 3 t_\mathrm{eddy}$ at
different resolutions. The dynamo completion time $t_\mathrm{sat}$ is
numerically converged at $\approx 5 t_\mathrm{eddy}$ by $256^3$. The
best-fit kinematic growth time follows a power law in the resolution,
$\tau_1 \propto N^{-0.6}$. This is consistent with the value of $-2/3$
expected if the dynamo time is controlled by shearing at the smallest
scale, the cascade is Kolmogorov (i.e. $u_\ell \propto \ell^{1/3}$), and
the viscous cutoff $\ell_\nu$ occurs at a fixed number of grid zones. In
that case, $\tau_1 \sim t_\nu \propto \ell_\nu^{2/3} \sim N^{-2/3}$.

\subsection{Power spectrum of kinetic and magnetic energy}
The time development of kinetic and magnetic energy power spectra has
been studied for a single run with resolution $512^3$. We present
three-dimensional, spherically integrated power spectra with the
dimensions of $\rm{ergs} / \rm{cm}^3 / \rm{m}^{-1}$, defined as
\begin{subequations}
  \begin{align}
    P_K(k_i) &= \frac{1}{\Delta k_i}\sum_{\vsp{k} \in \Delta k_i}
    {\left|\mathcal{F}_{\vsp{k}}\left[\vsp{v}\sqrt{\rho/2}\right]\right|^2}\\
    P_M(k_i) &= \frac{1}{\Delta k_i}\sum_{\vsp{k} \in \Delta k_i}
    {\left|\mathcal{F}_{\vsp{k}}\left[\vsp{B}/\sqrt{8\pi}\right]\right|^2}
  \end{align}\label{eqn:pspec}
\end{subequations}
where the Newtonian versions of kinetic and magnetic energy are
appropriate since the conditions are only mildly relativistic. The
definitions in Equations \ref{eqn:pspec} satisfy $\int{P(k) dk} =
\langle E \rangle$ for $P_K$ and $P_M$. Figure
\ref{fig:pspec-time-devel-res} shows the power spectrum of kinetic and
magnetic energy at various times throughout the growth and saturation of
magnetic field. During the kinematic phase, the kinetic energy has a
power spectrum $P_K(k) \propto k^{-5/3}$ consistent with the Kolmogorov
theory for incompressible hydrodynamical turbulence, while $P_M(k,t)
\propto e^{t/\tau_1} k^{3/2}$ consistent with Kazenstev's
model. $P_M(k)$ maintains the same shape, but exponentiates in amplitude
at the time scale $\tau_1$ which is controlled by shearing at the
resistive scale. According to Kazentsev's model, $P_M(k)$ should peak at
the resistive scale. This is consistent with the observed peak in the
magnetic energy at roughly 10 grid zones, the same scale at which we
observe the viscous cutoff. This is also consistent with $\Pm=1$
expected from the numerical scheme employed.

When the magnetic energy at the resistive scale surpasses the level of
the kinetic energy at that scale, $P_M(k)$ changes shape. The
equipartition scale $\ell_{K,M}$ moves into the inertial range, and
migrates to larger scale until full saturation occurs with $\ell_{K,M}
\approx L/4$. The movement of $\ell_{K,M}$ to larger scale is associated
with the formation of coherent and dynamically substantial magnetic
structures of increasing size. The time-dependence has been suggested to
be $\ell_{K,M} \propto t^{3/2}$ \citep{Beresnyak:2009p5056}. In the
fully saturated state, the magnetic field is in scale-by-scale
super-kinetic equipartition throughout the inertial range, with $P_M(k)
\approx 4 P_K(k)$. The largest scales remain kinetically dominated so
that the numerically converged saturation level is $E_M \approx 0.6
E_K$.

\section{Discussion}
In this Letter we have determined the time scale and saturation level of
the small-scale turbulent dynamo operating in the conditions of binary
neutron star mergers. We have presented numerically converged
simulations showing that magnetic fields are amplified to the $\sim
10^{16}$G level within a small fraction of the merger dynamical time
($t_\mathrm{sat} \lesssim t_\mathrm{merge}$), independently of the seed
field strength. If hydrodynamical instabilities create fluctuating
velocities on the order of $0.1c$ as indicated by global simulations,
then each $1 \mathrm{km}^3$ turbulent volume dissipates $\gtrsim
10^{46}$ erg of magnetic energy per 0.1 ms. If $f_\mathrm{turb}$
represents the fraction of the merger remnant that contains such
turbulence, the magnetic energy dissipated during the merger is at least
\begin{equation}
  E_\mathrm{diss} \sim 10^{49} \mathrm{erg} \times
  \left( \frac{f_\mathrm{turb}}{10^{-2}} \right)
  \left( \frac{t_\mathrm{merge}}{1 \mathrm{ms}} \right)
\end{equation}

A fraction of that dissipation will occur through magnetic reconnection
in optically thin surface layers, supplying relativistic electrons which
synchrotron radiate in the merger remnant magnetosphere. If 5\% of that
magnetic energy dissipation creates radiation in the 15-150 keV band,
then the fluence at 200 Mpc would be $10^{-7} \rm{erg}/\rm{cm^2}$,
potentially rendering most merging neutron stars in the advanced LIGO
and Virgo detection volumes detectable by \emph{Swift} BAT. If so, then
merging neutron stars are accompanied by a prompt electromagnetic
counterpart, independently of whether a later merger phase yields a
collimated outflow capable of powering a short gamma-ray burst.

We suggest that merger flares may be present in the current sample of
short GRBs and may be roughly isotropic on the sky since they are seen
to distances where the cosmological matter distribution becomes
homogeneous. Searches for merger flares should seek to identify short
flares, not unlike soft-gamma repeaters, among the short burst
population. If mergers also produce short GRBs short-hard GRBs, then
merger flares may constitute a precursor component of the emission.

The presence of strong magnetic fields may also aid in the ejection of
neutron-rich material from surface layers of the merger remnant,
possibly enhancing the enrichment of inter-stellar medium by r-process
nuclei \citep{Freiburghaus:1999p121, Rosswog:1999p499}. Enhanced
production of r-process nuclei also increases the likelihood of EM
detection by radio-active decay powered afterglows, or ``kilonovae''
\citep{Li:1998p5004, Metzger:2010p5000}.

Finally, it has been shown that magnetic fields will significantly
influence the gravitational wave signature and remnant disk mass, if
they exist at the $10^{17}$G level \citep{Etienne:2012p5005}. Such
strong fields are unlikely in older neutron star binaries, but our
results suggest they may be revived, albeit at small scales, during the
merger. To have significant influence, those fields would have to fill a
considerable fraction of the merger volume. As we have shown in this
Letter, the overall magnetic energy budget is controlled by the
prevalence ($f_\mathrm{turb}$) and vigor ($\varepsilon_K$) of the
turbulent volumes. This fact motivates the use of higher resolution
global simulations aimed at measuring $f_\mathrm{turb}$ and
$\varepsilon_K$.

\acknowledgments

This research was supported in part by the NSF through grant
AST-1009863 and by NASA through grant NNX10AF62G issued through the
Astrophysics Theory Program. Resources supporting this work were
provided by the NASA High-End Computing (HEC) Program through the NASA
Advanced Supercomputing (NAS) Division at Ames Research Center.

\end{document}